\documentclass{PoS}
\usepackage{url}
\usepackage{graphicx}
\usepackage{epsfig}
\usepackage{psfrag}
\def\spose#1{\hbox to 0pt{#1\hss}}
\def\ltapprox{\mathrel{\spose{\lower 3pt\hbox{$\mathchar"218$}}
 \raise 2.0pt\hbox{$\mathchar"13C$}}}
\def\gtapprox{\mathrel{\spose{\lower 3pt\hbox{$\mathchar"218$}}
 \raise 2.0pt\hbox{$\mathchar"13E$}}}
\def\inapprox{\mathrel{\spose{\lower 3pt\hbox{$\mathchar"218$}}
 \raise 2.0pt\hbox{$\mathchar"232$}}}

\newcommand{\bea}{\begin{eqnarray}}
\newcommand{\eea}{\end{eqnarray}}

                                {\end{table}}

\title{Light quark physics from lattice QCD}

\ShortTitle{Light quark physics}

\author{\speaker{Jack Laiho}\\
        SUPA, Department of Physics and Astronomy, University of Glasgow, Glasgow, Scotland, UK\\
        E-mail: \email{jlaiho@physics.gla.ac.uk}}
        
\abstract{I review lattice calculations of quantities that involve light quarks, including light quark masses, the vector form factor $f_+(0)$ needed for semileptonic kaon decays, and kaon mixing.  Results for most of these quantities are now available from multiple groups.  Averages of these results are presented, along with a discussion of the methodology behind the averaging procedure.  Recent progress in calculations of $K\to\pi\pi$ matrix elements is also reviewed. 
}

\FullConference{The XXVIII International Symposium on Lattice Field Theory, Lattice2010\\
		June 14-19, 2010\\
		Villasimius, Italy}

\begin{document}

\section{Introduction}

Lattice calculations of light-quark quantities such as quark masses, the kaon bag parameter, and semi-leptonic form-factors play a crucial role in extracting the fundamental parameters of the Standard Model and in constraining new physics beyond it.  More complicated quantities such as non-leptonic kaon decays are very sensitive to new physics, but are more difficult to calculate using lattice QCD.  Nonetheless, great progress has been made on this front in just the last year.

Almost all of the more straightforward quantities mentioned above have been calculated by a number of different groups, and there is excellent agreement between them when we restrict ourselves to calculations for which a complete systematic error budget has been provided.  Furthermore, for almost all of the quantities considered here there is more than one result with precision comparable to the highest quoted precision for that quantity, so that this agreement is a nontrivial and important test of lattice methods.  This is true of all the light quark masses, including the strange quark mass, where it appears that the discrepancies of the past have now been largely resolved.  (See, e.g. Ref.~\cite{Colangelo:2010et} for a compilation of older quark mass results.)
  
In this talk I cover light quark masses ($u$, $d$, and $s$), the $K\to\pi\ell\nu$ form factor and the determination of $|V_{us}|$, the kaon bag parameter $B_K$, kaon bag parameters relevant for beyond the Standard Model physics, and $K\to\pi\pi$ matrix elements.  There are fully documented results from more than one group for most of these; when this is the case world averages are presented.  The averages presented here is work done in collaboration with Enrico Lunghi and Ruth Van de Water, where we update and extend to new quantities the averages presented in reference \cite{Laiho:2009eu}.  The latest averages (including quantities outside the scope of this review) can be found on our website www.latticeaverages.org.  After reviewing the procedure we adopted for the averages presented in this talk, I present status updates of the various quantities in turn, including the new averages.  

Related reviews presented at this conference include a review of the light pseudoscalar decay constants by Christian H\"olbling \cite{Hoelbling:2011kk}, and a review of the impact on phenomenology of many of the above quantities by Chris Sachrajda \cite{Sachrajda:2011tg}.

\section{Lattice averages}\label{averages}

Our main criteria for including a result in the world averages quoted here is that the calculation include a complete error budget, including all sources of systematic error, and also that the work be documented in either a publication or proceedings.  Thus, we do not include numbers that appear only in the slides of a conference talk.  For the most part, we restrict ourselves to including only $N_f=2+1$ flavor lattice results in the averages because $N_f=0$ and $2$ flavor calculations typically do not quote an error due to quenching, as this is notoriously difficult to estimate reliably.  Since we only include results where estimates of all relevant systematic errors have been made, we exclude most of the $N_f=2$ flavor results from the averages, though I show the most recent state-of-the-art two flavor calculations in the comparison plots.  We make an exception in the averages for the $N_f=2$ calculation of the $K\to\pi\ell\nu$ form factor at zero-recoil, $f_+(0)$, since chiral perturbation theory power counting can give a reasonable estimate for the size of the effect of quenching the strange quark for this quantity, and this error was included as part of a complete systematic error budget by the ETM Collaboration in ref~\cite{Lubicz:2009ht}. 

There is an additional small systematic error in all of the $N_f=2+1$ flavor lattice results due to the neglect of the dynamical charm quark from the simulations.  This error is typically not included in current error budgets, though the effect is believed to be small.  Because the charm quark is relatively heavy compared to $\Lambda_{QCD}$, the size of the effects due to neglecting the charm quark in the sea can be estimated using heavy-quark effective theory (HQET).  The leading corrections are of order $\alpha_s[\Lambda/(2m_c)]$, where the two is a combinatoric factor appearing generically in HQET.  Using reasonable estimates for these quantities (e.g., $\alpha_s=0.33$, $\Lambda=500$ MeV, $m_c=1.2$ GeV) suggests that the size of charm loop effects is around $1\%$, although this effect is likely to be further suppressed in $SU(3)$-breaking ratios such as $f_K/f_\pi$.  Simulations with $N_f=2+1+1$ flavors are underway by the ETM Collaboration \cite{Baron:2010bv} and the MILC Collaboration \cite{Bazavov:2010ru} to address this issue.  Note that this HQET power counting estimate cannot be applied to the strange quark, and that the effects due to quenching the strange quark are more difficult to quantify without a direct comparison between $N_f=2$ and $2+1$ flavor simulations.  Such a direct comparison is possible, since the ETM Collaboration has now calculated many quantities using $N_f=2$ flavors with all systematic errors under control except that due to quenching the strange quark.  Good agreement is found with 2+1 flavor calculations at the quoted level of precision, suggesting that the effect of quenching the strange quark is small.  Nonetheless, the precision on the $2+1$ flavor averages is typically better than that of the 2 flavor calculations and is of the size where one might expect dynamical strange quark effects to become visible.  Thus, to be consistent with our criteria of including only results with estimates of all the relevant systematic errors, we do not include $N_f=2$ flavor calculations in our averages, expect for the special case of $f_+(0)$, as mentioned above and discussed further in Section V.

In order to avoid underestimating the errors in the averages, we need to take into account correlations between the different calculations.  Although the full correlation matrices between the different calculations do not exist and would be difficult to construct, we can still account for correlations using the following conservative assumptions:  when calculations use the same gauge field ensembles for the same quantity, we assume the statistical errors are $100\%$ correlated.  When a systematic error is at all correlated between different calculations, we assume the correlation in this error between calculations is $100\%$.  This treatment of systematic correlations is conservative in that it will lead to somewhat of an overestimate of the total error in the average, but we feel it is a reasonable assumption given the information available.  Finally, we also adopt the Particle Data Group (PDG) prescription for combining several measurements whose spread is wider than what one would expect given the quoted errors.  The error on such an average is rescaled by the square root of the minimum of the $\chi^2/$ per degree of freedom \cite{Nakamura:2010zzi}.

A snapshot of the simulation parameters is presented in Table~\ref{tab:params} to give a sense of the range of lattice actions, lattice spacings, and light pion masses being used in the most recent results.  For the staggered simulations, I have quoted the root-mean-square pion mass for the minimum pion mass in the sea, while for the valence sector I have quoted the taste-goldstone pion mass as the minimum pion mass.  This gives a sense of how light the valence and sea pion masses are compared to other simulations and is based on a more detailed look at staggered chiral perturbation theory formulas \cite{Aubin:2003mg, VandeWater:2005uq, Bae:2010ki} for the quantities considered in this review that were computed using staggered quarks.  For simulations performed with domain wall valence quarks on a staggered sea, I again quote the root-mean-square pion mass for the minimum pion mass in the sea, while for the valence sector I quote the mass of the valence pion made of two domain wall quarks.  This table reflects the parameters used in ongoing simulations, not necessarily the parameters appearing in the averages for all quantities presented below.

\begin{table}
\caption{Snapshot of parameter values being used in numerical simulations reviewed in this talk.  The first column is the group name; the second is the number of dynamical fermions; the third is the action; the fourth is the approximate range of lattice spacings; the fifth is the dimensionless product $m_\pi L$, which gives a measure of expected finite-size effects; and the last column is the approximate minimal pion mass simulated, in both the sea and valence sectors.}
\bigskip
\label{tab:params}
\begin{tabular}{cccccc}
\hline \hline
Group & $N_f$ & action & $a$(fm) & $m_\pi L$ & $m_\pi^{\rm min}$ (MeV) \\
& & & & & sea/val \\
\hline
ETMC \cite{Boucaud:2008xu} & 2 & Twisted Mass & $0.05$-$0.10$ fm & $\gg 1$ & $280/280$ \\
MILC \cite{Bazavov:2009bb} & 2+1 & (Asqtad) staggered & $0.045$-$0.12$ fm & $>4$ & $250/180$ \\
RBC/UKQCD \cite{Aoki:2010dy} & 2+1 & Domain Wall & $0.085$-$0.11$ fm &$>4$ & $290/210$ \\
JLQCD \cite{Fukaya:2009fh} & 2+1 & Overlap & $0.11$ fm  & $\geq 2.7$  & $310/310$ \\
PACS-CS \cite{Aoki:2009ix} & 2+1 & Clover & $0.09$ fm  & $\geq 2.0$ & $140/140$ \\
BMW \cite{Durr:2010aw} & 2+1 & Clover & $0.065$-$0.125$ fm & $\geq 4$ & $120/120$ \\
ALV \cite{Aubin:2009jh} & 2+1 & DW on MILC & $0.06$-$0.12$ fm & $> 3.5$ & $250/210$ \\
HPQCD \cite{Davies:2009ih} & 2+1 & HISQ on MILC & $0.045$-$0.15$ fm & $\geq 3.7$ & $360/310$ \\
 \hline
 \end{tabular}
\end{table}

\section{Quark masses}

The quark masses are fundamental parameters of the Standard Model, but due to confinement do not appear as free particles in nature.  One must tune the lattice quark masses so that one reproduces the experimentally known hadron spectrum, with three experimental inputs needed to determine three quark masses (and a fourth experimental input to fix the scale).  The bare lattice quark masses are then known in the regularization scheme defined by the lattice action of any particular calculation, but to compare to results using other lattice actions or to be used as input in continuum calculations, must be converted to a standard continuum regularization scheme like $\overline{MS}$.  The matching factors needed to convert to the $\overline{MS}$ scheme are short distance parameters, and can be computed perturbatively.  Direct perturbative matching between a lattice scheme and the $\overline{MS}$ scheme is a technical challenge, especially at two-loop order or beyond.  The convergence of lattice perturbation theory is typically poor unless tadpole-improvement and a renormalized coupling are used \cite{Lepage:1992xa}.  The convergence also tends to be better behaved when improved actions with fat gauge links are used \cite{DeGrand:2002va}, but this increases the difficulty of going to higher order, and few results beyond one-loop are known.  Non-pertubative matching to a regularization independent continuum scheme can be performed \cite{Martinelli:1994ty}, and then the matching to $\overline{MS}$ can be done in continuum perturbation theory, where it is easier to go to higher orders.  It is also possible to renormalize the masses completely non-perturbatively using the Schr\"{o}dinger functional scheme \cite{Luscher:1992an, Sint:2000vc}.

\begin{figure}
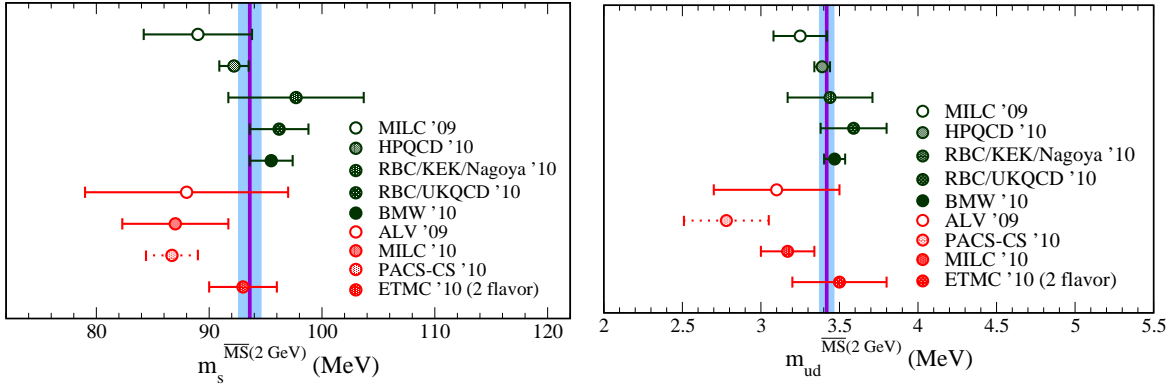

\begin{center}
\begin{picture}(147,70) 
\put(-1,0){\includegraphics[width=0.5\linewidth]{ms_proc_12_2010.eps}}
\put(78,1){\includegraphics[width=0.5\linewidth]{mud_proc_12_2010.eps}}
\end{picture}
\vspace{-1mm}
\caption{\emph{Left panel}: The strange quark mass $m_s^{\overline{MS}}(2 \textrm{GeV})$ from many groups, including the world average.  See the text for further details on color coding, etc.  \emph{Right panel}: The results for the average of the $u$ and $d$ quark masses $m_{ud}^{\overline{MS}}(2 \textrm{GeV})$ from many groups with world average.  Results in both figures are taken from Refs.~\cite{Bazavov:2009tw, Davies:2009ih, McNeile:2010ji, Blum:2010ym, Aoki:2010dy, Durr:2010vn, Laiho:2009, Bernard:2010, Aoki:2010wm, Blossier:2010cr}.  \label{fig:ms}}
\end{center}
\end{figure}

The latest results for the strange quark mass and for the averaged up and down quark mass $m_{ud}$ are shown in Figure~\ref{fig:ms}.  The notation in these and all subsequent average plots is as follows: for quantities that have been presented with full error budgets in a paper or proceedings, the data point is shown in green, and is included in the average.  Quantities that are not included in the average are shown in red.  This can be because the quantity has not yet been fully documented, or because potentially significant systematic errors have not yet been estimated.  If the situation is the later, the quantity is distinguished by a dashed line in the error bar.  A slight exception to this rule is that two flavor results are displayed without a dashed line, to emphasize that all errors except that due to quenching the strange quark have been included in the error estimate, but they are not included in the averages in most cases as discussed above and are labeled explicitly.  

The agreement between the different results for quark masses is striking, especially given the long history of disagreement between various $m_s$ determinations.  The only noticeable outlier in Figure~\ref{fig:ms} is the PACS-CS result \cite{Aoki:2010wm}, though this result does not yet include a full systematic error budget by the admission of the authors, so that the significance of the difference cannot yet be assessed.  The PACS-CS result uses a fully non-perturbative renormalization of $Z_m$ from the Schr\"{o}dinger functional method, and values of the quark masses down to the physical masses.  These simulations still involve rather small volumes ($m_\pi L\sim 2$) and there is so far only a single lattice spacing.  The quoted systematic errors include statistical errors, a systematic error due to reweighting to the physical quark masses, and a complete systematic error for $Z_m$.  The error budget does not yet include errors due to discretization effects and finite volume effects, but calculations to address the remaining systematic errors are in progress \cite{Taniguchi:2010pm}.

\begin{figure}
\begin{center}
\includegraphics[scale=1.0]{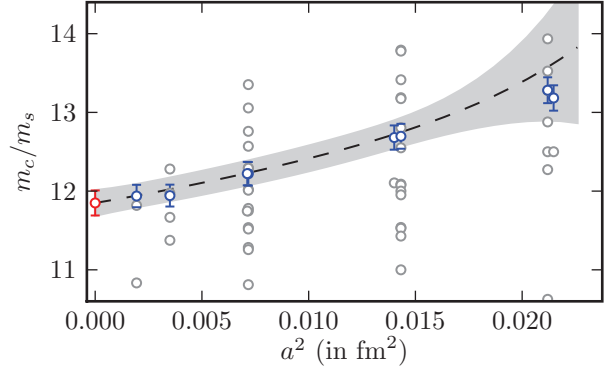}
\vspace{-1mm}
\caption{Continuum extrapolation of ratio $m_c/m_s$ from HPQCD \cite{Davies:2009ih}.\label{fig:HPQCD}}
\end{center}
\end{figure}

A few results in the quark mass determinations stand out due to their especially small errors.  The quark mass determinations of HPQCD use a novel method and have the smallest quoted errors \cite{Davies:2009ih}.  The HPQCD Collaboration make use of their determination of the charm quark mass using current-current correlators \cite{McNeile:2010ji, Allison:2008xk}, and the ratio of charm to strange quark masses \cite{Davies:2009ih} using a highly improved staggered quark (HISQ) action \cite{Follana:2006rc} to simulate a relativistic charm quark so that the charm and strange quark can be treated in the same formulation.  The ratio $m_c/m_s$, which receives no renormalization if both quarks are simulated in the same relativistic formalism, allows HPQCD to propagate their precise charm quark mass determination \cite{McNeile:2010ji} (using the HISQ formalism and four-loop continuum perturbation theory \cite{Chetyrkin:2006xg}) down to the light quark masses.  The continuum extrapolation of this quantity from HPQCD is shown in Fig.~\ref{fig:HPQCD}.  The $m_c/m_s$ ratio has also been calculated by the ETM Collaboration using two dynamical flavors \cite{Blossier:2010cr}, and it is in good agreement with the result of HPQCD.  The BMW Collaboration uses smeared clover quarks down to the physical light quark masses and the Rome-Southampton non-perturbative renormalization (NPR) method \cite{Martinelli:1994ty} and several lattice spacings down to $\sim0.05$ fm to control the perturbative matching \cite{Durr:2010vn}.  The RBC/UKQCD Collaborations use domain wall quarks and the Rome-Southampton NPR method, but they match to a non-exceptional momentum scheme \cite{Aoki:2010dy, Sturm:2009kb}, reducing the contamination from infrared effects and significantly reducing the size of the perturbative corrections compared to schemes that involve exceptional momenta.

\begin{figure}
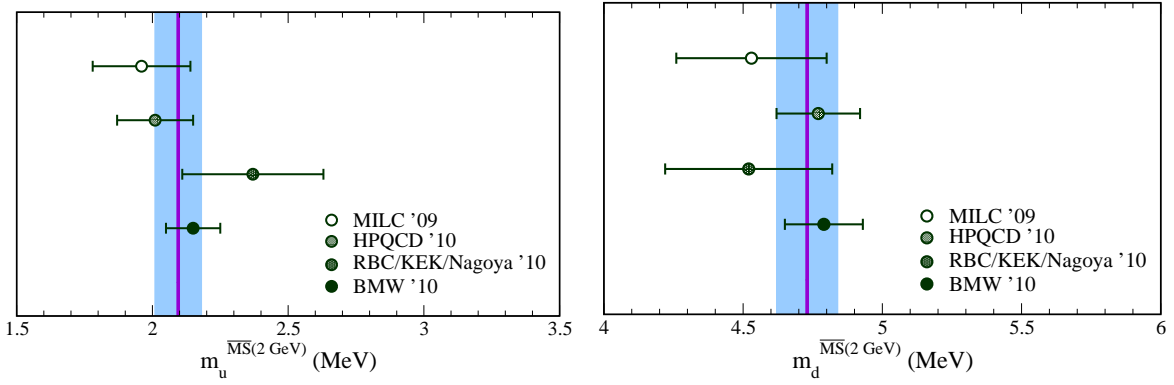

\begin{center}
\begin{picture}(147,70) 
\put(-1,0){\includegraphics[width=0.5\linewidth]{mu_12_2010.eps}}
\put(78,0){\includegraphics[width=0.5\linewidth]{md_12_2010.eps}}
\end{picture}
\vspace{-1mm}
\caption{\emph{Left panel}: The $u$ quark mass $m_u^{\overline{MS}}(2 \textrm{GeV})$ from different groups, including the world average.  \emph{Right panel}: The $d$ quark mass $m_{d}^{\overline{MS}}(2 \textrm{GeV})$ from different groups with world average.  This figure quotes results from Refs.~\cite{Bazavov:2009tw, Davies:2009ih, Blum:2010ym, Durr:2010vn}. \label{fig:mu_and_md}}
\end{center}
\end{figure}

Although the average of the $u$ and $d$ quark masses can be obtained in a fairly straightforward way from lattice simulations, the difference between the $u$ and $d$ quark masses requires an understanding of electromagnetic effects.  These have been incorporated into lattice calculations using different approaches.  The MILC results for $m_u$ and $m_d$ use the differences between the masses of charged and neutral pions and kaons, along with continuum estimates of the violation of Dashen's theorem to estimate the difference between the $u$ and $d$ quark masses \cite{Aubin:2004fs}.  HPQCD makes use of the MILC values of the ratios between the $u$ and $d$ quark masses, thus using the same input for electromagnetic contributions as MILC.  RBC/KEK/Nagoya include (quenched) QED explicitly in the lattice calculations, thus bypassing the need for continuum information on electromagnetic effects \cite{Blum:2010ym}.  They make use of two volumes to study finite volume effects, and they use $SU(2)$ heavy kaon chiral perturbation theory \cite{Allton:2008pn} to perform the chiral extrapolation, including electromagnetic effects.  BMW obtain information on the electromagnetic corrections from dispersive studies of $\eta\rightarrow 3\pi$ decays \cite{Durr:2010aw}.  Again, the agreement is impressive, as can be seen in Figure~\ref{fig:mu_and_md}, and the errors on the world averages presented here for the $u$ and $d$ quark masses are at the $5\%$ and $2\%$ level, respectively.  Note that the $u$ quark mass is over 20$\sigma$ from zero, so that the vanishing of this mass appears to be ruled out as a solution to the strong CP problem \cite{'tHooft:1976up, Jackiw:1976pf, Callan:1976je, Kaplan:1986ru}.

\begin{figure}
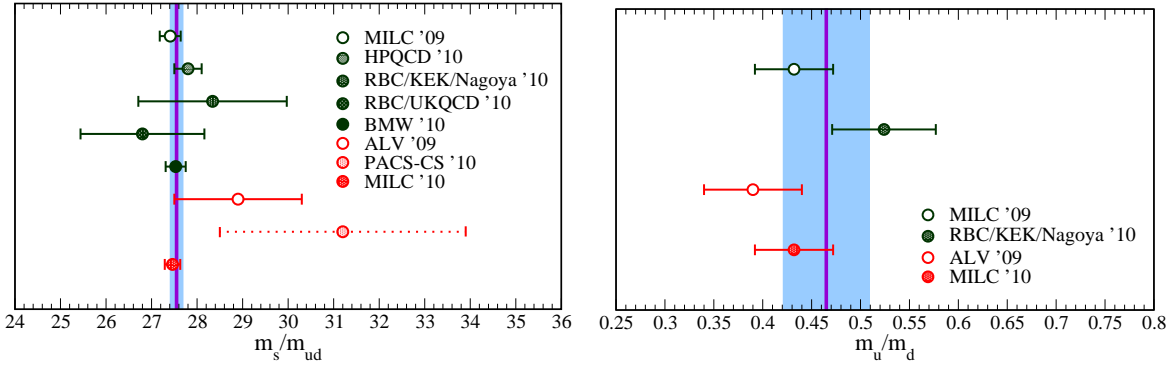

\begin{center}
\begin{picture}(147,70) 
\put(-1,0){\includegraphics[width=0.5\linewidth]{ms_mud_proc_12_2010.eps}}
\put(78,0){\includegraphics[width=0.5\linewidth]{mu_md_proc_12_2010.eps}}
\end{picture}
\vspace{-1mm}
\caption{\emph{Left panel}: The ratio of quark masses $m_s/m_{ud}$ from different groups, including the world average.  \emph{Right panel}: The ratio of quark masses $m_u/m_d$ from different groups with world average.  Results in these figures are taken from Refs.~\cite{Bazavov:2009tw, Davies:2009ih, McNeile:2010ji, Blum:2010ym, Aoki:2010dy, Durr:2010vn, Laiho:2009, Bernard:2010, Aoki:2010wm, Blossier:2010cr}.  \label{fig:mass_ratios}}
\end{center}
\end{figure}

Figure~\ref{fig:mass_ratios} shows the quark mass ratios $m_s/m_{ud}$ and $m_u/m_d$.  As noted above, these ratios are interesting to examine because the renormalization factor needed to quote the quark masses in a common scheme cancels.  For $m_s/m_{ud}$ there is, once again, impressive agreement between the various determinations, in particular those of MILC \cite{Bazavov:2009tw}, HPQCD \cite{Davies:2009ih}, and BMW \cite{Durr:2010vn}, where the total errors from each group are at the sub-percent level.  This precision agreement is noteworthy in part because it is not possible to get much guidance from experiment on the value of this ratio, unlike say the light hadron spectrum or pion and kaon decay constants, and no other non-perturbative methods quote errors that are competitive with the precision of the lattice results.  Leading order chiral perturbation theory ($\chi$PT) predicts a value around 26.0 \cite{Weinberg:1977hb}, but it is difficult to improve this result systematically to the same precision using $\chi$PT because of uncertainties in the low energy constants appearing at higher order in the chiral expansion.

\section{$K\rightarrow \pi\ell\nu$ semileptonic decay}

The semileptonic $K\to\pi\ell\nu$ decay can be used to obtain the CKM matrix element $V_{us}$ from the experimental branching fraction using \cite{Antonelli:2010yf}
\bea\label{eq:kl3} \Gamma_{K\ell 3}=\frac{G_F^2m_K^5}{192\pi^2}C_K^2S_{EW}(|V_{us}|f_{+})^2I_{K\ell}(1+\delta_{EM}^{K\ell}+\delta_{SU(2)}^{K\pi})^2,
\eea
where $S_{EW}=1.0232(3)$ is the short-distance electroweak correction, $C_K$ is a Clebsch-Gordan coefficeint, $f_+(0)$ is the form factor at zero momentum transfer, and $I_{K\ell}$ is a phase-space integral that is sensitive to the momentum dependence of the form factors.  The quantities $\delta^{K\ell}_{EM}$ and $\delta_{SU(2)}^{K\pi}$ are long-distance EM corrections and isospin corrections, respectively.
The value
\bea  |V_{us}|f_+(0) = 0.2163(5) 
\eea
has been determined from experimental measurements of $K\to\pi\ell\nu$ decays and non-lattice theory for the other inputs to Eq.~(\ref{eq:kl3})  \cite{Antonelli:2010yf}.  The non-perturbative information is encoded in the form factor $f_+(0)$, and once this is known from lattice QCD, a value for $V_{us}$ can be determined.  The value of $f_+(0)$ is already rather well constrained by $SU(3)$ chiral perturbation theory, an expansion in powers of $m_K^2/(8\pi^2f_\pi^2)$ \cite{Gasser:1984gg}.  One can write $f_+(0)=1+f_2+f_4+...$, where the first term is equal to one due to current conservation in the $SU(3)$ limit.  The correction $f_2$ does not contain any new unknown low energy constants, as required by the Ademollo-Gatto theorem \cite{Ademollo:1964sr}, and is predicted in terms of pion and kaon masses and the pion decay constant to be $f_2=-0.0226$.  We need to know $f_4$, if we are to do better, but this requires the determination of new higher order unknown low energy constants.  The value for $f_4$ was estimated by Leutwyler and Roos in 1984 \cite{Leutwyler:1984je} using a quark model; they obtained $f_4=-0.016(8)$, which gives $f_+(0)=0.961(8)$.

\begin{figure}[t]
\begin{center}
\includegraphics[scale=.5]{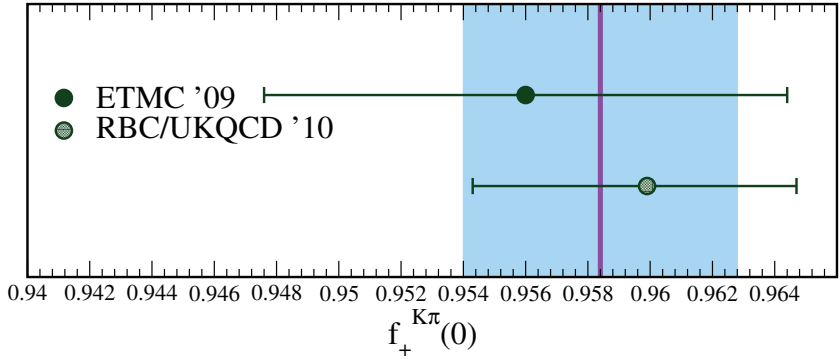}
\caption{Results for $f_+(0)$ including our average.  The results are quoted from Refs.~\cite{Lubicz:2009ht, Boyle:2010bh}.  \label{fig:Kl3}}
\end{center}
\end{figure}

There are only two lattice calculations of $f_+(0)$ that have complete systematic error budgets, one from ETMC from 2009 \cite{Lubicz:2009ht} and a more recent update in 2010 \cite{Boyle:2010bh} of an older calculation from RBC/UKQCD \cite{Boyle:2007qe}, with the average shown in Figure~\ref{fig:Kl3}.    As mentioned in Section~\ref{averages}, we include the ETMC result in our world average even though it is with $N_f=2$ flavors because there is a credible estimate of the systematic error due to quenching the strange quark.  The ETMC result has already been reviewed by Vittorio Lubicz \cite{Lubicz:2010nx} in his lattice review talk from last year, but it is worth recalling the treatment of the error due to quenching the strange quark for $f_+(0)$.  In the chiral effective theory the $N_f$ dependence can be incorporated; this modifies the chiral logarithms and changes the low energy constants from their $N_f=3$ to $N_f=2$ values.  Although it is straightforward to use the known chiral logarithm expressions to adjust for unphysical quenching effects, the $N_f$ dependence of the low energy constants is not known.  For $f_+(0)$  the effect of using $N_f=2$ sea quarks can be corrected for in the $f_2$ term exactly, since there are no new low energy constants appearing to this order, and the strange sea quark dependence is a known expression involving chiral logarithms.  The $f_4$ term requires the knowledge of new low energy constants, and these will take their $N_f=2$ values in the ETMC calculation.  The remaining error in $f_+(0)$ due to the difference between $f_4$ in the $N_f=2$ and 3 flavor theories is estimated by looking at the difference between $f_4$ in the $N_f=2$ and $N_f=0$ calculations.  The full difference between $f_4$ in the 2 flavor theory and the 0 flavor (completely quenched) theory is taken as the error due to quenching the strange quark for this quantity \cite{Lubicz:2009ht}.

The RBC/UKQCD result for $f_+(0)$ \cite{Boyle:2010bh} is an update of their previous result \cite{Boyle:2007qe} from 2007.  This new calculation improves upon the previous one by using twisted boundary conditions \cite{Sachrajda:2004mi, Bedaque:2004ax}  to remove the need to interpolate in $q^2$ to get the form factor at zero-recoil.  They also apply a different choice of chiral extrapolation, where analytic next-to-next-to leading order (NNLO) terms that do not obey the Ademollo-Gatto theorem are included.  Such terms are possible if $f_\pi$ is used in the NLO expression instead of $f_0$, the decay constant in the $SU(3)$ limit.  This is because when reordering the series to use $f_\pi$ in the NLO expression, analytic terms that do not respect the mass interchange symmetry can appear at NNLO.  RBC/UKQCD quote $f_+(0)=0.9599(34)(^{+31}_{-43})(14)$, where the first error is statistical, the second is due to the chiral extrapolation, and the third is an estimate of discretization errors \cite{Boyle:2010bh}.

There were progress updates for $f_+(0)$ from the FNAL/MILC Collaborations using staggered quarks \cite{Bailey:2010vz} and from the JLQCD Collaboration using overlap quarks \cite{JLQCD:2010ru}.  The FNAL/MILC calculation is using the method developed by HPQCD for $D\to K\ell\nu$ \cite{Na:2011hy} to get a result for $f_+(0)$ using
\bea f_+(0)=f_0(0)=\frac{m_s-m_q}{m_K^2-m_\pi^2}\langle \pi|S|K\rangle_{q^2=0}
\eea
so that no renormalization is required.  This avoids the use of non-local vector currents and does not require multiple three-point correlators to form various double ratios.  Twisted boundary conditions are used to calculate $f_0$ at the $q^2=0$ point.  The disadvantage of this method is that one cannot obtain $f_0(q^2)$ for $q^2\neq 0$, but this is still sufficient to determine $|V_{us}|$, since the shape dependence of the form factor is usually taken from experiment.  

The JLQCD calculation was done at a single lattice spacing and a somewhat small (1.7 fm) volume.  The $q^2$ dependence was modeled using various functional forms to interpolate to $q^2=0$, and the shape dependence was in reasonable agreement with experiment.  Work with twisted boundary conditions and larger volumes is in progress \cite{JLQCD:2010ru}.

\section{Kaon mixing}

The constraint on the unitarity triangle coming from kaon mixing can be expressed as
\bea  |\epsilon_K|=C_\epsilon \kappa_\epsilon B_K A^2\overline{\eta}\{-\eta_1 S_0(x_c)(1-\lambda^2/2)+\eta_3 S_0(x_c,x_t) + \eta_2 S_0(x_t)A^2\lambda^2(1-\overline{\rho}) \},
\eea
where $C_\epsilon$ is a collection of experimentally determined parameters, $\kappa_\epsilon$ represents long-distance contributions and a correction due to the fact that the quantity $\phi_\epsilon \neq 45$ degrees \cite{Buras:2008nn}, $B_K$ is the kaon bag parameter, the $\eta_i S_0$ are perturbative coefficients, and $\lambda$, $A$, $\overline{\rho}$, $\overline{\eta}$ are CKM matrix elements in Wolfenstein parameterization.  The experimental determination of $|\epsilon_K|$ leads to a constraint on the unitarity triangle in the shape of a hyperbolic band in the $\overline{\rho}$-$\overline{\eta}$ plane.  The main non-perturbative input needed from the lattice to implement this constraint on the CKM unitarity triangle is the kaon bag parameter $B_K$.

\begin{figure}
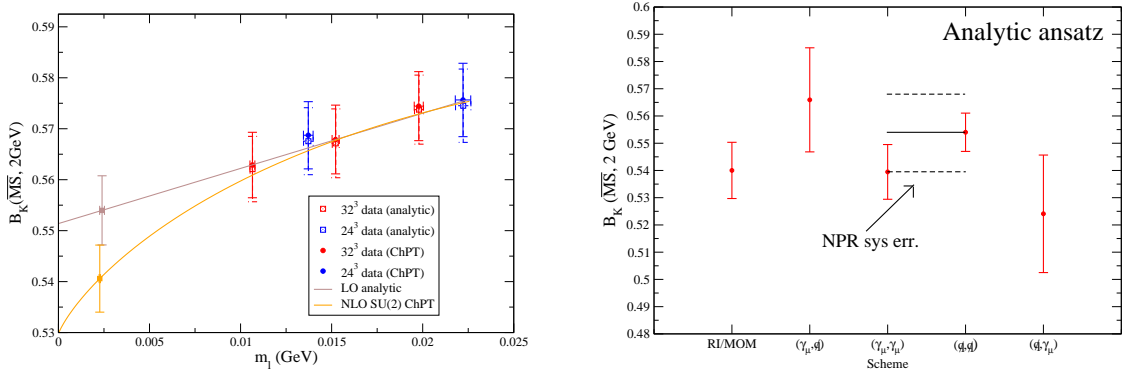

\begin{center}
\begin{picture}(147,70) 
\put(-1,0){\includegraphics[width=0.46\linewidth]{bk_unitary_flavour_su2_comparison.eps}}
\put(78,0){\includegraphics[width=0.46\linewidth]{plot_analytic.eps}}
\end{picture}
\vspace{-1mm}
\caption{\emph{Left panel}: Extrapolation in light quark mass for $B_K$ from RBC/UKQCD \cite{Aoki:2010pe}.  \emph{Right panel}: Comparison of $B_K$ in different renormalization schemes from RBC/UKQCD \cite{Aoki:2010pe}.\label{fig:BK_RBC}}
\end{center}
\end{figure}

There have been three recent updates on this quantity: from the RBC/UKQCD Collaborations \cite{Aoki:2010pe}, from the SBW Collaboration \cite{Bae:2010ki}, and from ETMC \cite{Constantinou:2010qv}.  The RBC/UKQCD Collaborations have updated their eariler result \cite{Antonio:2007pb} using domain-wall quarks with a number of improvements, including a second lattice spacing to allow a continuum limit to be taken, and the use of various non-perturbative renormalization schemes with non-exceptional momentum to perform the matching to the continuum \cite{Lehner:2011fz}.  The RBC/UKQCD results have also modified their approach to the chiral extrapolation.  For their central value, they average the result using $SU(2)$ heavy-kaon $\chi$PT to perform the extrapolation in light quark mass and the result using a simple linear extrapolation.  This approach was motivated by the absence of detectable curvature in their data, and the tendency of their $SU(2)$ fit to undershoot $f_\pi$.  Figure~\ref{fig:BK_RBC} shows the light-quark mass extrapolation for $B_K$ for both the linear and the $SU(2)$ fits.  RBC/UKQCD used multiple RI-SMOM schemes with non-exceptional momenta to determine the matching factor for $B_K$.  Figure~\ref{fig:BK_RBC} shows the comparison of results between the various RI-SMOM schemes, as well as results from the RI/MOM scheme.  The systematic error associated with the NPR matching is also shown in Fig~\ref{fig:BK_RBC}.  The final result quoted for $\hat{B}_K$ is 0.749(7)(21)(3)(15), where the errors are statistical, chiral extrapolation, finite volume, and renormalization \cite{Aoki:2010pe}.

The SBW Collaboration has adopted a mixed-action approach, using HYP-smeared staggered quarks \cite{Bae:2008qe} on the MILC asqtad ensembles, with four lattice spacings down to 0.045 fm \cite{Bae:2010ki, Bae:2010bq}.  The matching is done to one-loop order in lattice perturbation theory \cite{Kim:2011pz}.  The chiral and continuum extrapolation is done using $SU(2)$ staggered chiral perturbation theory \cite{Yoon:2010bm}.  The $SU(2)$ formulation has much simpler expressions than the $SU(3)$ case, where many new parameters specific to the staggered formalism enter \cite{VandeWater:2005uq}.  SBW quote as their main result a value $\hat{B}_K=0.724(12)(43)$, where the errors are statistical and the sum of systematic errors in quadrature \cite{Bae:2010ki}.  The dominant error is currently due to the one-loop perturbative matching.  Non-perturbative matching is in progress.

The ETM Collaboration has a result for $B_K$ using a mixed action \cite{Frezzotti:2004wz} with three lattice spacings down to 0.07 fm \cite{Constantinou:2010qv}.  The valence action is Osterwalder-Seiler \cite{Osterwalder:1977pc} and the sea sector is that of the ETMC $N_f=2$ twisted mass ensembles \cite{Boucaud:2008xu}.  They quote a value of $\hat{B}_K=0.733(29)(16)$, where the first error contains statistical, chiral extrapolation/fit, and matching errors, while the second contains an error due to different assumptions of ${\cal O}(a^2p^2)$ dependence in the RI-MOM scheme matching factor \cite{Constantinou:2010qv}.  A calculation by ETMC using 2+1+1 flavors is in progress.
 
\begin{figure}[t]
\begin{center}
\includegraphics[scale=.4]{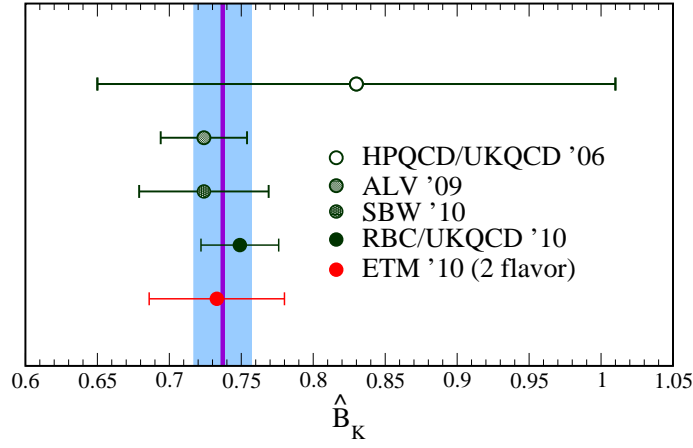}
\caption{Results for $\hat{B}_K$ including the world average.  Results are taken from Refs.~\cite{Gamiz:2006sq, Aubin:2009jh, Bae:2010ki, Aoki:2010pe, Constantinou:2010qv}.  \label{fig:BK}}
\end{center}
\end{figure}

 The world average for $\hat{B}_K$ is shown in Fig.~\ref{fig:BK}.  All of the results are in good agreement, which is impressive given the different discretizations and methods employed in the various calculations.  The effect of quenching the strange quark appears to be rather small, as we can see by comparing the ETMC result with that of the 2+1 flavor average.
 
 The ETM Collaboration has presented the first preliminary unquenched (two-flavor) results for four-quark operators that contribute to kaon mixing in the presence of new physics \cite{Dimopoulos:2010wq}.  The effective Hamiltonian relevant for beyond the Standard Model physics contains several new four-quark operators, in addition to the one associated with $B_K$.  Only four new matrix elements are required, however, since for the new operators only the parity-even parts are needed because the strong interaction conserves parity.  
 
 The bag parameters associated with the new operators are defined by \cite{Dimopoulos:2010wq}
 \bea \langle \overline{K}^0 |{\cal O}_1(\mu)| K^0 \rangle = B_K(\mu)\frac{8}{3}m^2_K f^2_K,
 \eea
 \bea  \langle \overline{K}^0 |{\cal O}_i(\mu)| K^0 \rangle = C_i B_i(\mu)\left[\frac{m^2_K f_K}{m_s(\mu) + m_d(\mu)} \right]^2,
 \eea
 where $C_i=\{-5/3, 1/3, 2, 2/3\}$, $i=2,...,5$, when we take the basis of operators used in Ref.~\cite{Dimopoulos:2010wq}.  The matrix element of ${\cal O}_1$ is just the Standard Model contribution to kaon mixing.  Note that the other operators do not vanish in the chiral limit.  

The ETMC calculation includes three lattice spacings, fairly light pion masses (down to $\sim 280$ MeV) and non-perturbative renormalization using the RI-MOM scheme.  They find that the chiral extrapolation is not very sensitive to the choice of fit function, and we show the results for the $B_i$ parameters from the quadratic fit of Ref.~\cite{Dimopoulos:2010wq} in Table~\ref{tab:BSM_Bi}.  The errors are statistical only.

\begin{table}
\begin{center}
\caption{Preliminary values of BSM kaon bag parameters in the $\overline{MS}$ scheme with two dynamical flavors from ETMC \cite{Dimopoulos:2010wq}.  Errors are statistical only.\label{tab:BSM_Bi}} 
\vspace{-1mm}
\begin{tabular}{cc} \\ \hline\hline
$i$ & $B_i$   \\[0.5mm] \hline
2 & 0.56(0.04)  \\
3 & 1.43(0.13) \\
4 & 0.76(0.06) \\
5 & 0.63(0.09) \\
\hline\hline
\end{tabular}
\end{center}\end{table}

\section{$K\to \pi\pi$}

Lattice calculations of non-leptonic $K\to\pi\pi$ decays are challenging because of the two-hadron final state, but they are important for phenomenology.  Lattice calculations of $K\to\pi\pi$ matrix elements have the potential to give us a first principles determination of the decades old $\Delta I=1/2$ rule, and would finally allow us to use the experimental measurement of $\varepsilon'/\varepsilon$ as a precision constraint on the Standard Model \cite{Antonelli:2009ws}.  The Standard Model prediction for $\varepsilon'/\varepsilon$ is 
\bea  \textrm{Re}\left(\frac{\varepsilon_K'}{\varepsilon_K}\right) \approx \frac{\omega}{\sqrt{2}|\varepsilon_K|}\left[\frac{\textrm{Im}(A_2)}{\textrm{Re}(A_2)}-\frac{\textrm{Im}(A_0)}{\textrm{Re}(A_0)}\right],
\eea
where $A_0$ and $A_2$ are the amplitudes for $K\to\pi\pi$ decays into definite isospin states, and real and imaginary refer only to the part of the amplitude that becomes complex due to the presence of the weak phase.  The smallness of the parameter $\omega=\textrm{Re}(A_2)/\textrm{Re}(A_0)\approx 0.05$ is a manifestation of the $\Delta I=1/2$ rule.

$K\to\pi\pi$ matrix elements are difficult to calculate on the lattice because the Maiani-Testa no-go theorem \cite{Maiani:1990ca} tells us that we cannot extract physical matrix elements from Euclidean correlation functions with multi-hadronic final states.  Due to the restriction of working in Euclidean time, the most straightforward lattice implementation of calculating $K\to\pi\pi$ matrix elements only works if the final state pions are at rest, or at some other set of unphysical kinematics.  Two general strategies have emerged for getting around this problem.  One strategy is  to construct $K\to\pi\pi$ matrix elements indirectly using the low energy constants (LEC's) of chiral perturbation theory as determined from simpler lattice matrix elements such as $K\to 0$ and $K\to\pi$ \cite{Bernard:1985wf}.  It was shown in Refs.~\cite{Boucaud:2001mg, Laiho:2002jq, Lin:2002nq, Laiho:2003uy} that all LEC's through next-to-leading order could be obtained from relatively simple lattice quantities.  However, this method has the disadvantage that the convergence of $SU(3)$ chiral perturbation theory at the physical kaon mass is slow, and it is not clear whether $K\to\pi\pi$ matrix elements can be computed in this way to a useful precision \cite{Li:2008kc, Laiho:2010ir}.  

A method for calculating $K\to\pi\pi$ matrix elements directly at physical kinematics was introduced by Lellouch and L\"uscher \cite{Lellouch:2000pv}.  The Lellouch-L\"uscher method exploits the finite lattice volume to obtain the matrix elements directly by tuning the volume so that the first excited state of the two pion state matches the kaon mass.  The direct method is straightforward to implement, though it is computationally demanding because it requires large lattice volumes ($\sim 6$ fm) and physical light quark masses.  Improvements to the method have been introduced so that the non-zero momentum pion state becomes the ground state and smaller volumes can be used \cite{Kim:2002np, Christ:2005gi, Kim:2005zzb, Christ:2009ev}.  The RBC/UKQCD Collaborations have made significant progress using the direct method, with preliminary results at nearly the physical quark masses and physical kinematics for the $\Delta I=3/2$ decay channel \cite{Goode:2011kb}; this is discussed below.

\begin{figure}
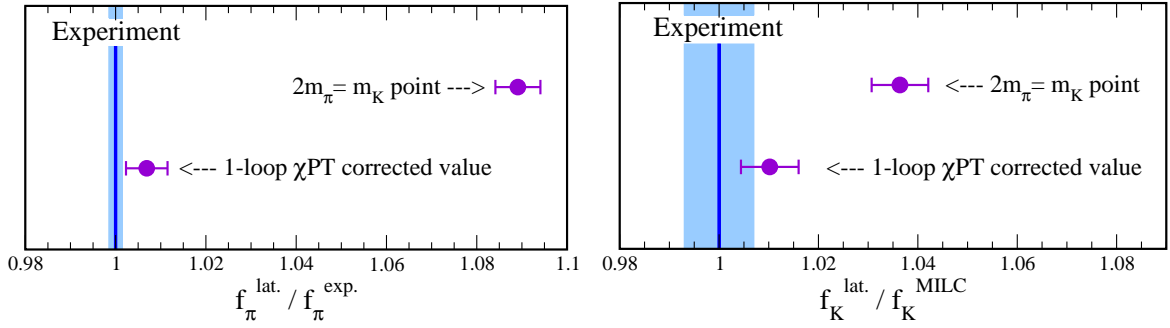

\begin{center}
\begin{picture}(147,70) 
\put(-1,1){\includegraphics[width=0.505\linewidth]{fpi_2pi_ratio.eps}}
\put(78,1){\includegraphics[width=0.5\linewidth]{fK_2pi_ratio.eps}}
\end{picture}
\vspace{-1mm}
\caption{Demonstration of the size of NLO $SU(3)$ corrections to quantities evaluated at $2m_\pi=m_K^{\rm phys}$ for $f_\pi$ (left) and $f_K$ (right).  Errors on the circular points are statistical only.  Uncertainties in the vertical $f_\pi$ and $f_K$ error bands include both statistical and systematic errors \cite{Laiho:2010ir}. \label{fig:2pi}}
\end{center}
\end{figure}

A technique for improving upon the indirect method was presented by the author and Van de Water, where we exploited the fact that one can simulate $K\to\pi\pi$ matrix elements with the pions produced at rest \cite{Laiho:2010ir}.  If one takes the pion mass to be $1/2 m_K$, this amplitude can be computed directly, since it is not forbidden by the Maiani-Testa theorem, a fact known for quite some time \cite{Dawson:1997ic}.  Thus, we tune the light quark masses such that $m_K=m_K^{\rm phys}$ and $m_\pi= m_K^{\rm phys}/2$.  One can then correct for the unphysical kinematics using fixed order $SU(3)$ $\chi$PT, where the low energy constants can be obtained from simpler quantities, like $K\to\pi$.  Because the kaon is tuned to its physical value, the terms involving kaons (etas) are correct (nearly correct) to all orders in the $SU(3)$ chiral expansion.  Thus the $10$-$30\%$ precision of NLO $SU(3)$ $\chi$PT now appears in a small correction factor, rather than the entire amplitude.  This can be tested for known quantities like $f_K$ and $f_\pi$, results of which are shown in Fig.~\ref{fig:2pi}.  These plots illustrate the fact that the NLO corrections to $f_\pi$ are below $10\%$ for this method, and the corrections to $f_K$ are below $5\%$.  Also, the one-loop corrections account for most of the difference between the $2m_\pi=m_K$ values and the accepted values.

\begin{figure}
\begin{center}
\includegraphics[scale=0.5]{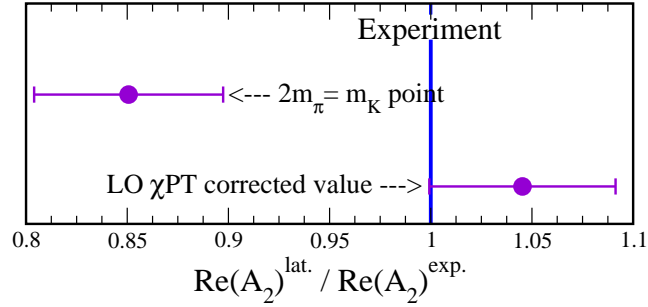}
\vspace{-1mm}
\caption{The correction to $\textrm{Re}(A_2)$ at $2m_\pi=m_K^{\rm phys}$ using leading order $SU(3)$ $\chi$PT compared to experiment \cite{Laiho:2010ir}. \label{fig:2piKpipi}}
\end{center}
\end{figure}

The value for Re$(A_2)$ at physical kinematics from JL and Van de Water is calculated similarly, where the value at the $2m_\pi=m_K$ point is corrected using the leading order $\chi$PT expression,
\bea  \langle \pi^+ \pi^-|O^{(27,1),(3/2)}|K^0 \rangle_{\rm LO} = \frac{4iB_0 f_0}{3}(m_K^2-m_\pi^2),
\eea
where $f_0$ and $B_0$ are the pion decay constant and $B_K$ in the $SU(3)$ chiral limits, respectively.  The leading order correction factor is then
\bea\label{eq:correct} \delta^{\rm LO}_{\chi \textrm{PT}} =[(m_K/2)^2-m_\pi^2]/[m_K^2-(m_K/2)^2],
\eea
which is only $23\%$.  This correction is shown in Fig.~\ref{fig:2piKpipi}, along with the experimental value.  This corresponds to a value of $\textrm{Re}(A_2)=1.568(86)\times 10^{-8}$ GeV, where the error is statistical only \cite{Laiho:2010ir}.  A preliminary estimate of the systematic errors in this approach are shown in Table~\ref{tab:total_err}.  The error is dominated by the $12\%$ $\chi$PT truncation error, though this may improve when the correction factor in Eq.~(\ref{eq:correct}) is known to one-loop in $\chi$PT.

\begin{table}
\begin{center}
\caption{Estimated total error budget for $\textrm{Re}(A_2)$ from JL and Van de Water \cite{Laiho:2010ir}. Each source of uncertainty is given as a percentage.\label{tab:total_err}} 
\vspace{-1mm}
\begin{tabular}{lr} \\ \hline\hline
uncertainty & $\qquad \textrm{Re}(A_2)$   \\[0.5mm] \hline
statistics & 4.7\%  \\
$\chi$PT truncation error & $12\%$ \\
uncertainty in leading-order LECs & 4\% \\
discretization errors & 4\% \\
finite volume errors & few percent  \\
renormalization factor & 3.4\%  \\
scale and quark-mass uncertainties & 3\%  \\
Wilson coefficients & few percent \\
\hline
total & less than 20\% \\
\hline\hline
\end{tabular}
\end{center}\end{table}

The calculation of $K\to\pi\pi$ decays from RBC/UKQCD uses the direct Lellouch-L\"uscher approach, and they have made significant progress, including a preliminary result for matrix elements in the $\Delta I=3/2$ channel with close to $10\%$ errors \cite{Goode:2011kb}.  These calculations are done using the new Dislocation Suppressing Determinant Ratio (DSDR) domain-wall quark ensembles being generated by RBC/UKQCD \cite{Jung:2010jt} with volumes of $32^3\times 64$ with $L_s=32$ at an inverse lattice spacing of around $1.4$ GeV.  This corresponds to a spatial box size of around 4.5 fm.
The lightest unitary pion mass is 180 MeV, and a lighter valence pion with mass around 140 MeV is used for the central value.  The main errors contributing to the RBC/UKQCD calculation of Re$(A_2)$ are given in Table~\ref{tab:total_errRBC}.  The largest error is the estimate of scaling violations due to the use of somewhat coarse lattices at a single lattice spacing and the fact that $K\to\pi\pi$ matrix elements scale as the lattice spacing cubed.  The RBC/UKQCD result for Im$(A_2)$ is expected to have a similar error once the nonperturbative renormalization is completed.

\begin{table}
\begin{center}
\caption{Estimated total error budget for $\textrm{Re}(A_2)$ from RBC/UKQCD \cite{Goode:2011kb}. Each source of uncertainty is given as a percentage.\label{tab:total_errRBC}} 
\vspace{-1mm}
\begin{tabular}{lr} \\ \hline\hline
uncertainty & $\qquad \textrm{Re}(A_2)$   \\[0.5mm] \hline
statistics & 5.8\%  \\
scaling violations & $8.5\%$ \\
finite volume effects & $7\%$ \\
partial quenching & $2\%$ \\
pion phase shift & $2\%$  \\
meson masses and 2-pion energies & $1.2\%$  \\
\hline
total & 11\% \\
\hline\hline
\end{tabular}
\end{center}\end{table}

\begin{figure}
    \begin{tabular}{cc}
	 \epsfysize=1.0in \rotatebox{0}{\epsffile{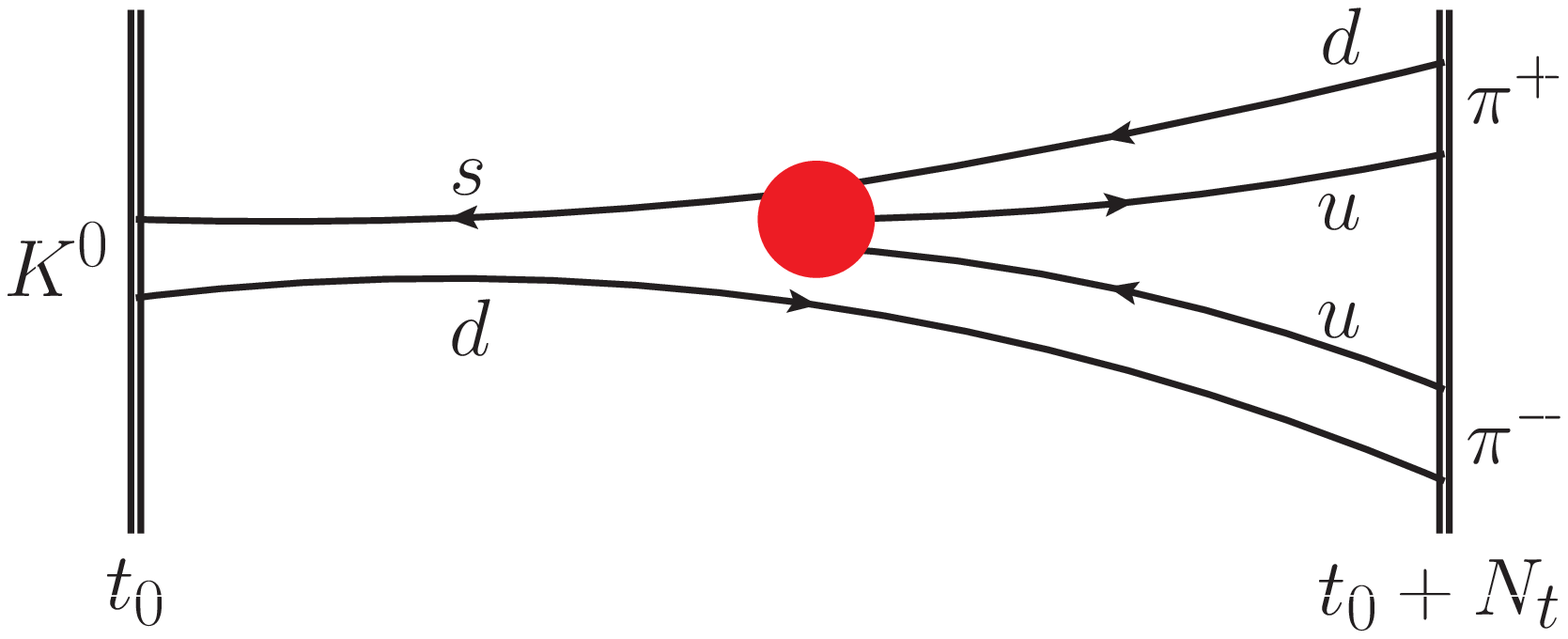}} &  \epsfysize=1.0in \rotatebox{0}{\epsffile{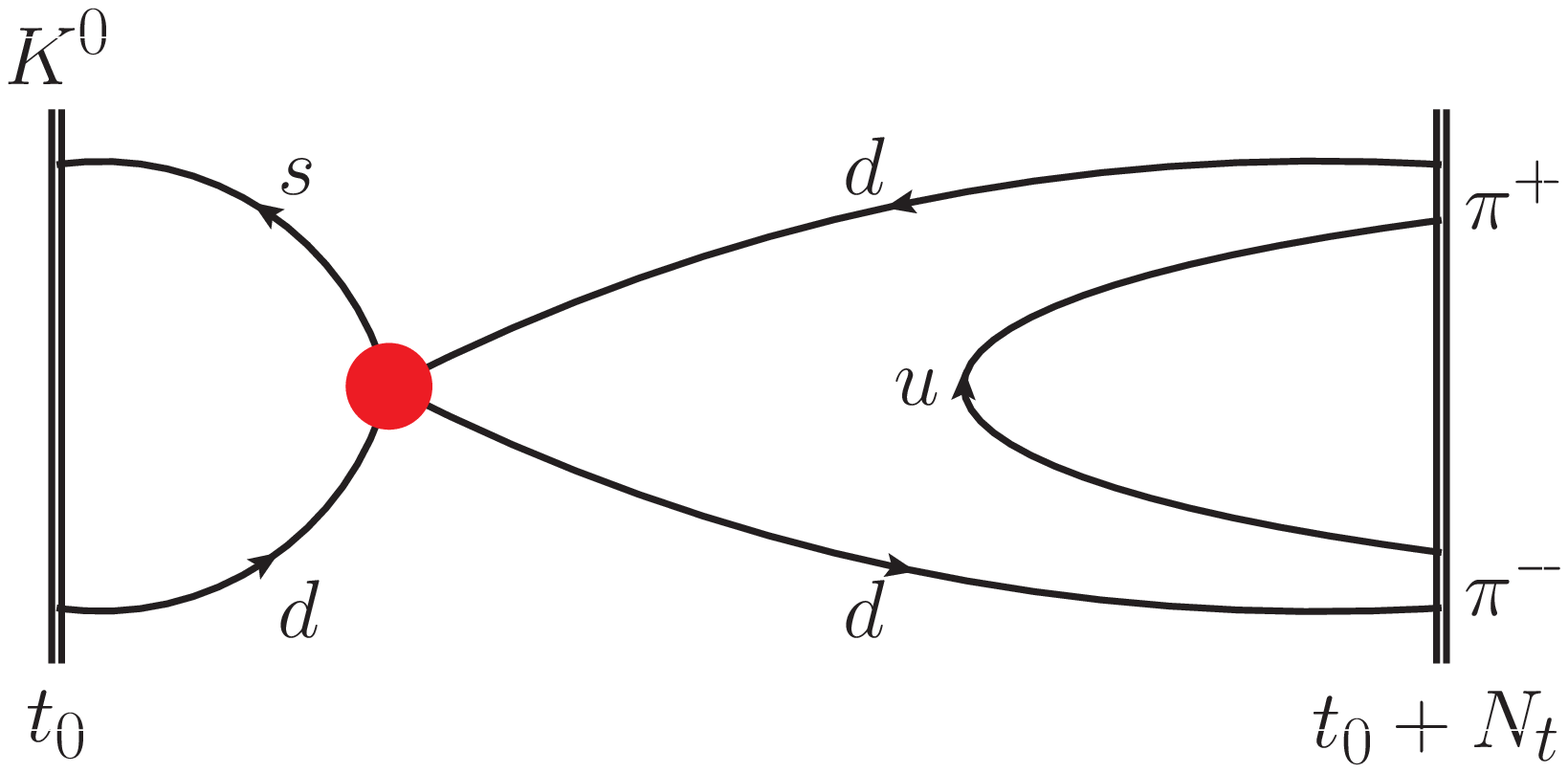}}  \\
	(a)  & (b)  \\\\
	\epsfysize=1.0in \rotatebox{0}{\epsffile{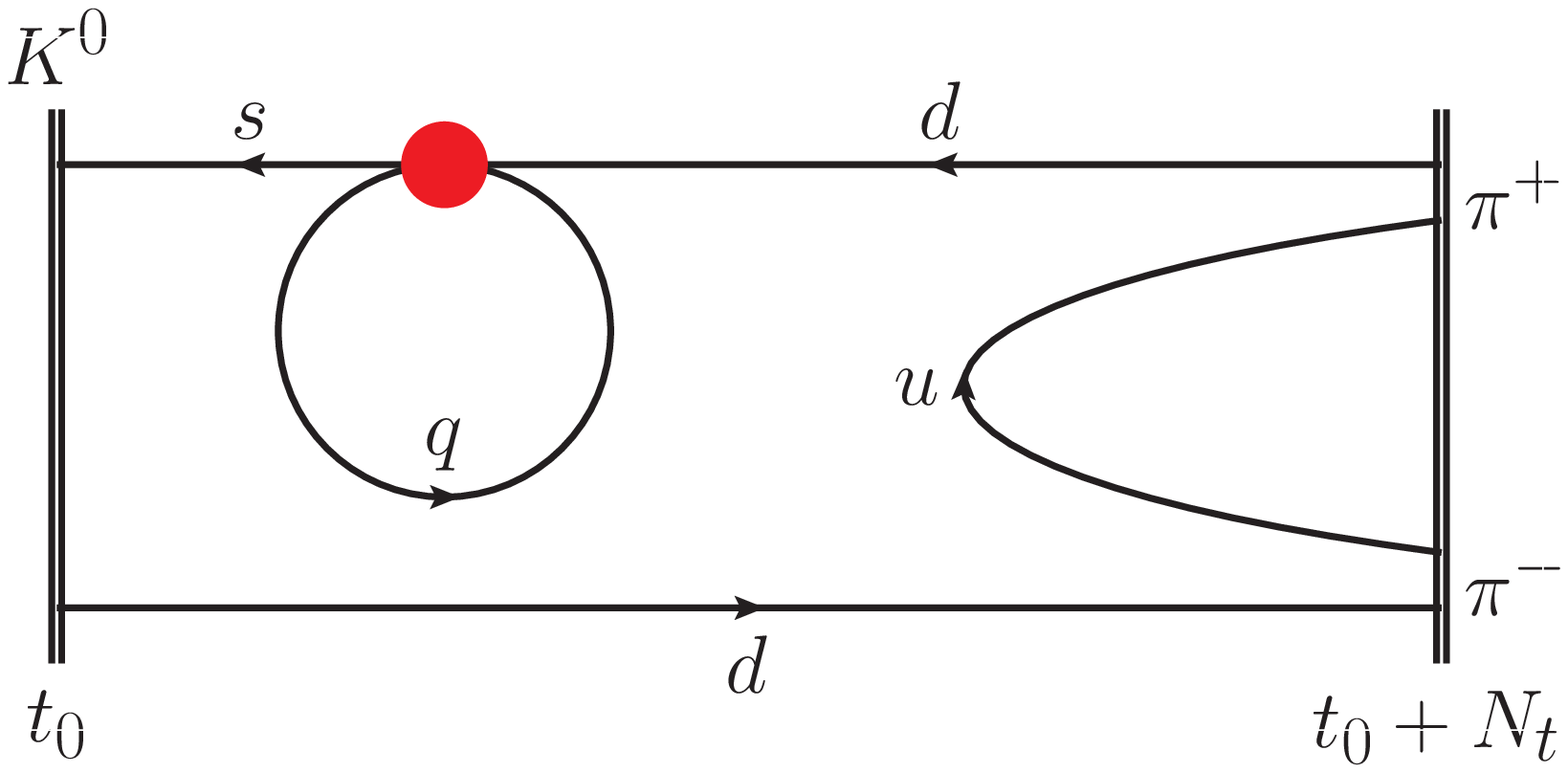}} &  \epsfysize=1.0in \rotatebox{0}{\epsffile{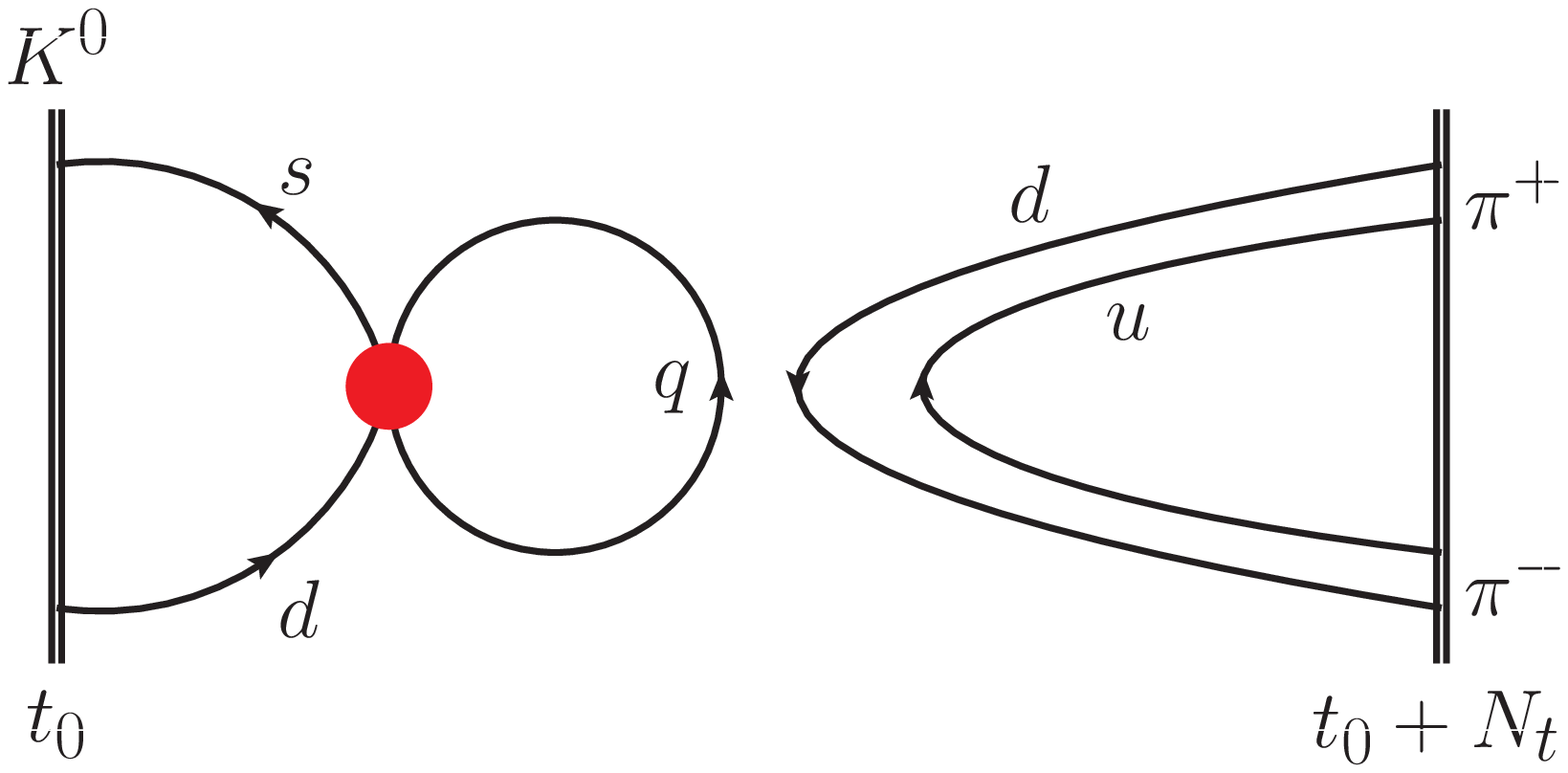}}  \\
	(c)  & (d)  \\
    \end{tabular}
    \caption{Quark flow diagrams for $K\to\pi\pi$ in the $\Delta I = 1/2$ channel.\label{fig:Kpipi_1_2}}
\end{figure}

The calculation of the $\Delta I=1/2$ rule is more difficult for a number of reasons.  One is the presence of power divergent contributions arising from mixing with lower dimensional operators.  This problem has been addressed by the use of chiral fermions, where the operator subtraction is straightforward \cite{Noaki:2001un, Blum:2001xb}.  Another problem is the presence of enhanced finite-volume effects that afflict the calculation when the light valence quark masses are not the same as in the sea \cite{Colangelo:1997ch, Lin:2003tn}.  This was an especially serious problem for quenched attempts to calculate $\Delta I=1/2$ kaon matrix elements \cite{Lin:2002nk}, but is under control when sea quarks of the correct mass are included in the calculation.  Another difficulty is the appearance of disconnected quark flow diagrams, leading to the need for very high statistics.  The contractions at the level of quark flow are shown in Fig.~\ref{fig:Kpipi_1_2} for the $\Delta I=1/2$ channel.  The red circle is the insertion of the four-quark operator.  Additional diagrams with a quark current insertion are not shown, but are needed to perform the power divergent operator subtraction.  Figure~\ref{fig:Kpipi_1_2}(d) shows the disconnected diagram that is problematic due to the need for high statistics.  This problem will likely be solved by more computing and better inversion algorithms.

\begin{figure}
\begin{center}
\begin{picture}(147,70) 
\put(-3,0){\includegraphics[width=0.5\linewidth]{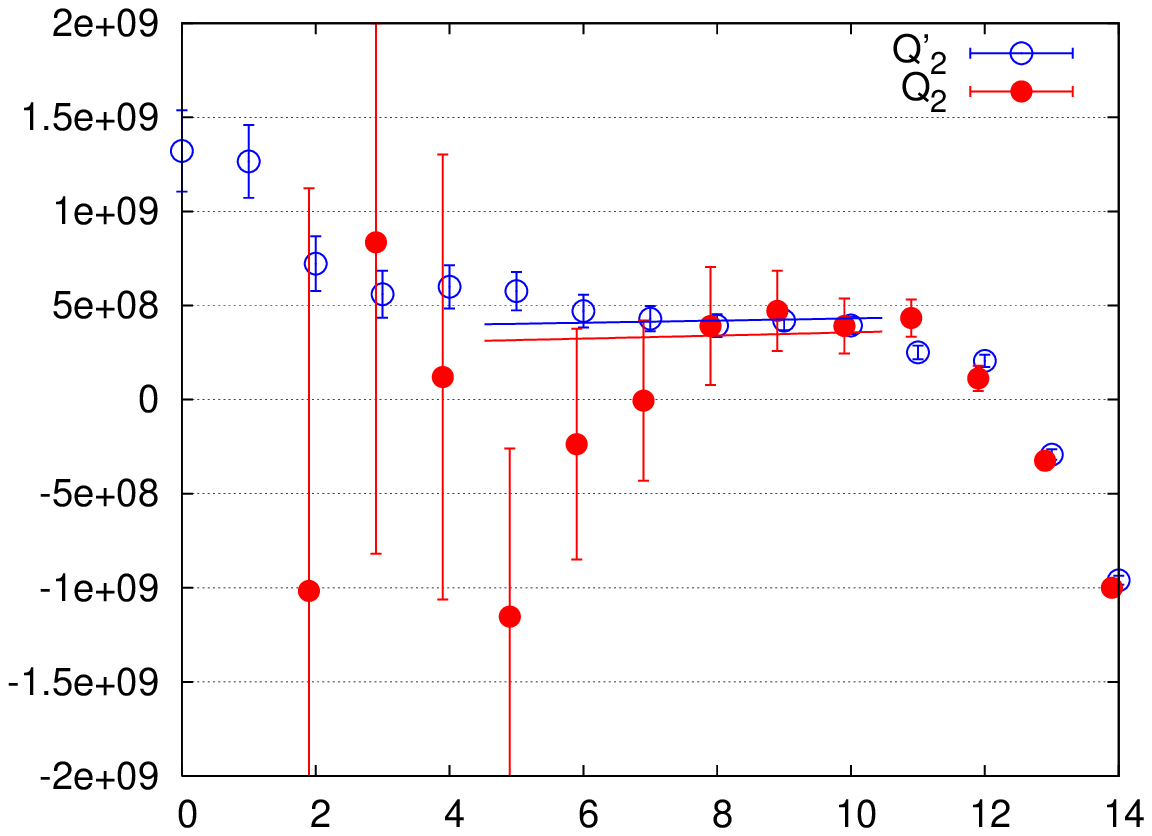}}
\put(75,0){\includegraphics[width=0.5\linewidth]{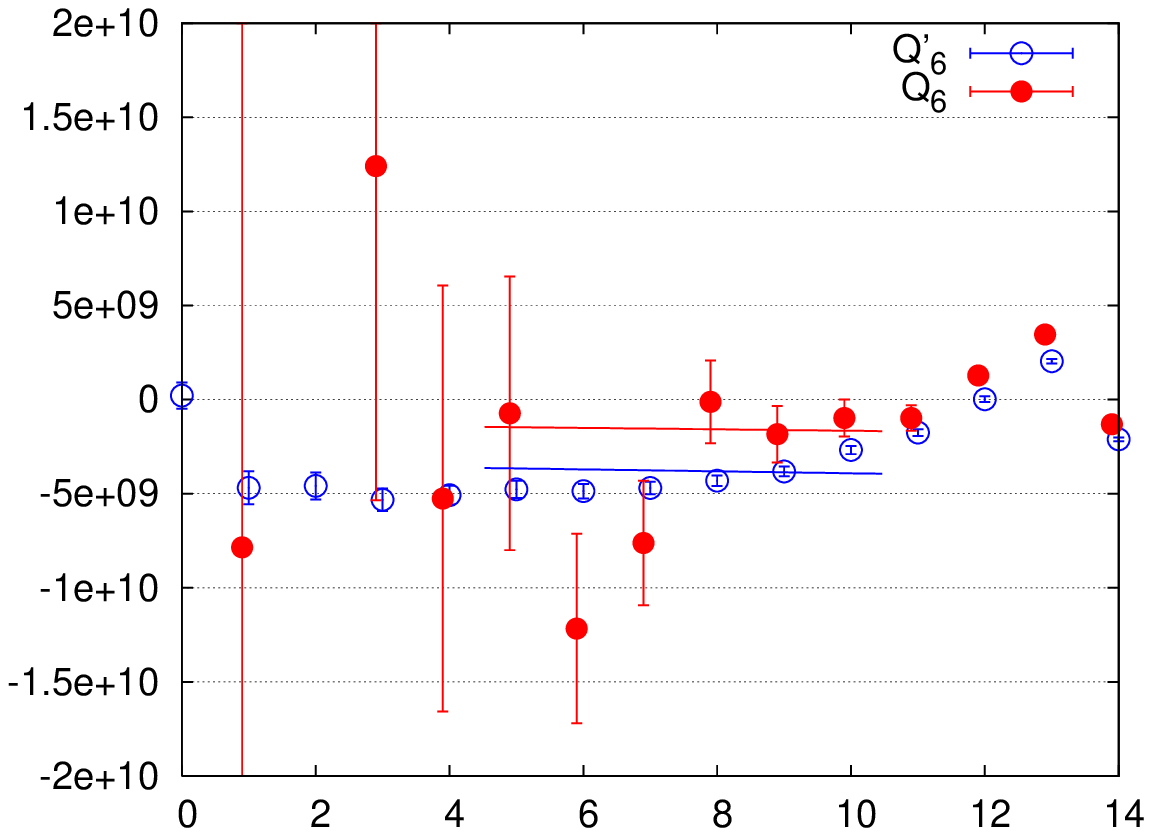}}
\end{picture}
\vspace{-1mm}
\caption{\emph{Left panel}: $K\to\pi\pi$ matrix elements at zero momentum for $Q_2$ with and without disconnected diagrams from RBC/UKQCD \cite{Liu:2010fb}. \emph{Right panel}: $K\to\pi\pi$ matrix elements at zero momentum for $Q_6$ with and without disconnected diagrams from RBC/UKQCD \cite{Liu:2010fb}.\label{fig:Kpipi_RBC}}
\end{center}
\end{figure}

Results for $K\to\pi\pi$ correlation functions in the $\Delta I=1/2$ channel from RBC/UKQCD were presented by Qi Liu \cite{Liu:2010fb}, and they are shown in Fig.~\ref{fig:Kpipi_RBC}.  These correlators were obtained from 2+1 flavor domain wall ensembles with Iwasaki gauge action generated by RBC/UKQCD with volume $16^3\times32$ and $L_s=16$, with pion masses around $420$ MeV.  Propagator inversions were performed on each time slice for 400 configurations.  The operator $Q_2$ gives the dominant contribution to Re$(A_0)$ at renormalization scales typical of lattice calculations, while $Q_6$ gives the dominant contribution to Im$(A_0)$.  The blue (open) circles show the contribution to the correlator from the diagrams (a), (b), and (c) in Fig.~\ref{fig:Kpipi_1_2}, while the red (filled) circles include all four diagrams, including the disconnected diagram Fig.~\ref{fig:Kpipi_1_2}(d).  The comparison between open and closed circles in Fig.~\ref{fig:Kpipi_RBC} shows the large statistical errors introduced by the disconnected diagram.  Table~\ref{tab:ReA0} also illustrates this point with entries for Re$(A_0)$ and Im$(A_0)$ with and without the disconnected diagram.  The error on Re$(A_0)$ is about $25\%$, while the error on Im$(A_0)$ indicates that more statistics are needed to be sure of a signal.  It was reported at the conference by Liu that for non-zero momentum the $\Delta I=1/2$ correlators barely had a signal, even without the disconnected diagrams.  Again, improvements are expected by going to larger lattices, as well as bigger machines and better inversion algorithms.

\begin{table}
\begin{tabular}{cccc}
\hline \hline
 Re$(A_0)_{\rm no \ discon}$ & Re$(A_0)$ & Im$(A_0)_{\rm no \ discon}$ & Im$(A_0)$ \\ \hline
 $38.7(2.1)\times 10^{-8}$ & $30(8)\times 10^{-8}$ & $-63.1(5.3)\times 10^{-12}$ & $-29(22)\times 10^{-12}$ \\
\hline
\end{tabular}
\caption{Results from RBC/UKQCD for on-shell $K\to\pi\pi$ matrix elements with $m_\pi=420$ MeV and zero momentum with and without disconnected contributions \cite{Liu:2010fb}.  Errors are statistical only.  \label{tab:ReA0}}
\end{table}

Finally, I observe that the indirect method for $K\to\pi\pi$ discussed above is not likely to match the precision possible with the direct method in the $\Delta I=3/2$ channel, but could be useful for the $\Delta I=1/2$ channel, given the significant amount of computing needed to reach non-zero momentum at the physical kinematics in this channel.

\section{Summary and Outlook}

Results for the simplest quantities in light quark physics are now in impressive agreement.  To take one example, many groups now have results for quark masses using dynamical ensembles, multiple lattice spacings, and improved techniques for computing the renormalization factor, and there is agreement at the few percent level, with systematic errors under control.  This is an important achievement for the lattice.  Other quantities that are important for flavor physics show similarly impressive agreement, and it has become clear that averages are necessary to maximize the impact of these lattice results on constraining new physics.  I have reviewed the approach to averaging that my collaborators E. Lunghi and R. Van de Water and I have adopted \cite{Laiho:2009eu}, and I have presented our updated averages for many quantities involving light quarks.  More difficult quantities like $K\to\pi\pi$ matrix elements in the $\Delta I=3/2$ channel are now within reach, and preliminary results with errors at the $10-20\%$ have been presented \cite{Goode:2011kb, Laiho:2010ir}.  The $\Delta I=1/2$ channel is more difficult but may be attainable in the next few years.

\acknowledgments

I wish to thank the organizers of ``Lattice 2010" for an excellent conference.  I also thank many colleagues for communicating their work to me in advance of my talk and for helpful discussions.  I thank Claude Bernard, Peter Boyle, Ting-Wai Chiu, Norman Christ, Christine Davies, Petros Dimopoulos, Elvira Gamiz, Taku Izubuchi, Karl Jansen, Takashi Kaneko, Jangho Kim, Weonjong Lee, Laurent Lellouch, Matthew Lightman, Qi Liu, Vittorio Lubicz, Antonin Portelli, Chris Sachrajda, Francesco Sanfilippo, Steve Sharpe, Amarjit Soni, Y. Taniguchi, Boram Yoon, and Taniguchi Yusuke. 

I also thank my collaborators Enrico Lunghi and Ruth Van de Water for all of the work that went into producing the averages presented here.

\end{document}